\documentclass[aps,letterpaper,showkeys,footinbib,twocolumn]{revtex4-1}

\usepackage{cancel}
\usepackage{amsthm}         
\usepackage{amsmath} 	   
\usepackage{amssymb}       
\usepackage{amsfonts}
\usepackage{graphicx} 	
\usepackage{color} 	
\usepackage{enumitem} 	
\usepackage{inputenc}
\usepackage{comment}
\usepackage[normalem]{ulem}
\usepackage{natbib}
\usepackage{footnote}
	
\begin{document}

\title{Implications of Planck2015 for inflationary, ekpyrotic and anamorphic bouncing cosmologies}

\author{Anna Ijjas}
\affiliation{Princeton Center for Theoretical Science, Princeton University, Princeton, NJ 08544, USA}
\author{Paul J. Steinhardt}
\affiliation{Department of Physics, Princeton University, Princeton, NJ 08544, USA}
\affiliation{Princeton Center for Theoretical Science, Princeton University, Princeton, NJ 08544, USA}

\date{\today}

\begin{abstract} 
The results from Planck2015, when combined with earlier observations from WMAP, ACT, SPT and other experiments, were the first observations to disfavor the ``classic" inflationary paradigm.     To satisfy the observational constraints,
 inflationary theorists have been forced to consider plateau-like inflaton potentials that introduce more parameters and more fine-tuning, 
problematic initial conditions, multiverse-unpredictability issues,  and   a new  `unlikeliness problem.'  Some propose turning instead to a ``postmodern" inflationary paradigm in which the cosmological properties in our observable universe are only locally valid and set randomly, with completely different properties (and perhaps even different physical laws) existing in most regions outside our horizon.    By contrast, the new results are consistent with the simplest versions of ekpyrotic cyclic models in which the universe is smoothed and flattened during a period of slow contraction followed by a bounce, and another promising bouncing theory, anamorphic  cosmology, has been proposed that can produce distinctive predictions.
\end{abstract}

\maketitle 

The Planck2015 \cite{Ade:2015lrj} and Planck2013 \cite{Ade:2013nlj} observations, combined with the results by the Wilkinson Microwave Anisotropy Probe (WMAP), Atacama Cosmology Telescope (ACT) and South Pole Telescope (SPT) teams have shown that
the measured spatial curvature is small; the spectrum of primordial density fluctuations is nearly scale-invariant; there is a small spectral tilt, consistent with a simple dynamical mechanism that caused the smoothing and flattening; and the fluctuations are nearly gaussian.  These features are all consistent with the simplest textbook inflationary models.  At the same time,  Planck2015 also confirmed that $r$, the ratio of the tensor perturbation amplitude to the scalar perturbation amplitude, is less than  0.1, a result that virtually eliminates all the simplest textbook inflationary models.  The development is notable because, as emphasized by Ijjas et al. \cite{Ijjas:2013vea},  it is the first time that the classic inflationary picture has been in conflict with observations.

The results have led theorists to consider alternative  plateau-like inflationary models whose parameters can be adjusted to reduce the expected value of $r$ while retaining all the rest of the classic predictions.  
However, as we explain in this brief review, the remaining models raise new issues. They require more parameters, more tuning of parameters, more tuning of initial conditions, a worsened multiverse-unpredictability problem, and a new challenge that we call the inflationary `unlikeliness problem.'    

One response to these problems has been that they should be ignored.  
The classic inflationary picture should be abandoned in favor of an alternative `postmodern' inflationary picture that allows enough flexibility to fit any combination of observations. The classic and postmodern inflationary pictures are so different that they ought to be viewed as distinct paradigms to be judged separately. 

A more promising response to Planck2015 has been to develop ``bouncing'' cosmologies in which the large-scale properties of the universe are set during a period of slow contraction and the big bang is replaced by a big bounce.  For example, a new, especially simple version of {\it ekpyrotic}  (cyclic) cosmology has been identified that fits all current observations, including nearly Gaussian fluctuations and small $r$ \cite{Li:2013hga,Fertig:2013kwa,Ijjas:2014fja,Levy:2015awa}.  Also, {\it anamorphic} bouncing cosmologies have been introduced that use yet different ways to smooth and flatten the universe during a contracting phase and generate a nearly scale-invariant spectrum of perturbations \cite{Ijjas:2015zma}.

We will first review the problems that Planck2015 imposes on classic inflation, the version that most observers consider.  We will briefly discuss the conceptual problems of initial conditions and multiverse that have been known and unresolved for decades.  Then we will turn to the tightening constraints resulting from Planck2015 and other recent experiments.  We will  review and critique postmodern inflation that some theorists now advocate.  Finally, we will turn to the promising new developments in bouncing cosmologies, both ekpyrotic and anamorphic approaches.
 
\section{Classic inflation}   

Most discussions of inflation focus on the inflaton potential ($V(\phi)$ where $\phi$ is the inflaton field), leaving the impression that this alone is needed to determine the predictions of an inflationary scenario.   In actuality, three independent inputs are needed:  the initial conditions, the inflaton potential, and the measure used to quantitatively compare the probabilities of different outcomes.  

The initial conditions are commonly specified  at the earliest time when classical general relativity begins to be a good approximation for describing cosmic evolution, typically the Planck time. Roughly, the {\it inflaton potential} determines a family of classical trajectories, some of which do and some of which do not include a long period of inflation; the initial conditions pick out a subset of trajectories; and the {\it measure} defines the relative ``weight'' among the subset of trajectories needed to compute the predictions.  Note that the measure may depend on factors other than the range of initial conditions, such as the number of $e$-folds of inflation.
 
The term ``classic inflation" used in this paper refers to the standard inflationary picture held by most observers and
described in most texts and popular descriptions.  The classic picture
aims for the simplest initial conditions, the simplest potentials and the simplest common-sense measure.   The notion is that, after the big bang, some regions of space have the properties required to undergo a period of accelerated expansion that smoothes and flattens the universe, leaving only tiny perturbations that act as sources of cosmic microwave background fluctuations and seeds for galaxy formation.  Although most regions of space emerging from the big bang may not have the correct conditions to start inflation, inflation exponentially stretches the volume of the regions that do have the right conditions.   Using volume-weighting as the measure, smooth and flat regions are assumed to dominate the universe by the end of inflation. For potentials with a minimum of fields (one) and a minimum of fine-tuning of parameters, there are \textit{generic} inflationary predictions: a spatially flat and homogeneous background universe with a nearly scale-invariant, red-tilted spectrum of primordial density fluctuations ($n_S \sim 0.94-0.97$), significant gravitational-wave signal ($r \sim 0.15-0.4$), and negligible non-gaussianity ($f_{\textsc{nl}} \sim 0$). 
 
\section{The classic problems with classic inflation}

Three fundamental problems afflict all classic inflationary models: (1) the fine-tuning problem; (2) the initial conditions problem; and (3) the multiverse problem.  Although all three have been known for more than three decades, no satisfactory solution has been found for any of them.  Rather, attempts to resolve the problems have shown that they are more pernicious and resistant than originally imagined. 

{\it Fine-tuning problem:}  All inflationary potentials require orders of magnitude of parameter fine-tuning to yield the observed amplitude of the primordial density fluctuations ($\delta\rho/\rho\sim 10^{-5}$).  For example,  the simplest inflaton potential, $V(\phi) = \lambda \phi^4$, has a single field and a single dimensionless parameter; to obtain the observed amplitude, $\lambda$ must be fine-tuned by 15 orders of magnitude compared to its natural value (order unity).  Although the tuning can take various forms, the same degree of tuning must be incorporated into any inflationary model.  No plausible way for avoiding fine-tuning has been identified.  Supersymmetry or shift symmetry have been invoked for explaining why the tuning is not spoiled by quantum corrections, but these symmetries do not explain why the value of $\delta\rho/\rho$ should be so small in the first place.  (Anthropic arguments fail since they favor larger values of $\delta\rho/\rho$ that produce exponentially more galaxies and planets.)  

Some contend that, the inflationary fine-tuning problem should not be regarded as  serious since it is comparable to the weak hierarchy problem of the standard model.  Our view is that both fine-tuning problems should be taken seriously; both are good motivation to seek a better theory; but the fine-tuning problem of inflation is especially troubling because inflation was introduced specifically to eliminate the fine-tuning problems of an earlier model, the original big bang picture. 

A second generic fine-tuning problem that is mentioned less often entails the spectral tilt.  It is often claimed that inflation predicts a nearly scale-invariant spectrum of scalar (curvature) perturbations, which corresponds to a small spectral tilt.   This claim is misleading.  The spectral tilt $1-n_S$ depends on the equation-of-state $\epsilon(N)$ and variation with $N$, the number of e-folds of inflation remaining:
\begin{equation}\label{ns}
1-n_S = 2 \epsilon - \frac{d \ln \epsilon}{dN},
\end{equation}
where $\epsilon = (3/2) (\rho+)/\rho$ for an ideal fluid with pressure $p$ and energy density $\rho$.  The condition for inflation (accelerated expansion) is $\epsilon<1$, which allows for substantial deviations from scale-invariance according to Eq.~(\ref{ns}). To have near scale-invariance consistent with observations, it is necessary that $\epsilon <0.03$, a small  fraction of the allowed region.  From this we can observe that near scale-invariance is not a natural prediction of inflation; instead, model-builders impose the condition on $\epsilon$ to match the observations.

Theorists (including us) have argued that a small value may be justified on the grounds that a simple scaling law, $\epsilon(N) =1/N$, where $N$ is the number of e-folds of inflation, is a natural choice for the equation of state \cite{Khoury:2003vb,Ijjas:2013sua}.   However, plausible the argument may have been, it can no longer be considered because it produces a a large value of $r$ that is inconsistent with Planck2015 observations.  More complicated expressions have been proposed with more parameters \cite{Mukhanov:2014uwa}, but, once that is allowed, the entire range $\epsilon <1$ is obtainable and the problem of explaining the small value of $\epsilon$ returns.

{\it Initial conditions problem:}
The probability of a region of space having the right initial conditions to have 60 or more $e$-folds of  inflation is exponentially small \cite{Penrose:1988mg,Gibbons:2006pa,Berezhiani:2015ola}.  Furthermore, given an inflaton potential,  there exist exponentially more homogeneous and flat cosmic solutions with little or no inflation than solutions with  60 or more $e$-folds of inflation \cite{Gibbons:2006pa}, according to a standard classical statistical mechanical reasoning based on a Liouville phase-space measure.  Some \cite{Schiffrin:2012zf} dispute the use of the Liouville measure, but other approaches give the same qualitative conclusion \cite{Berezhiani:2015ola}.  No matter which analysis is used, the problem traces back to the fact that the inflaton potential energy density must dominate for inflation to begin, but this energy density grows slowly compared to nearly all other forms of energy density (matter, radiation, gradient, curvature and inflaton kinetic energy density) when extrapolating back in time during the pre-inflationary phase.  Hence, it is very unlikely to have an initial condition after the big bang where the potential energy density dominates sufficient to have 60 e-folds  of inflation.

{\it Multiverse problem:}
The  multiverse  (sometimes called the measure problem)  is the consequence of  eternal inflation \cite{Steinhardt:1982kg,Vilenkin:1983xq}.  Assuming smooth, classical evolution of the inflaton field, inflation comes to an end in a finite amount of time according to when the inflaton reaches the bottom of its potential.  However, generically, classical evolution is sometimes punctuated by large quantum fluctuations, including ones that kick the inflaton field uphill, far from its expected classical course.  These regions undergo  additional  inflation that rapidly makes them occupiers of most of the volume of the universe.   Inflation then continues to amplify rare quantum fluctuations that keep space inflating, leading to eternal inflation.  Continuing along this line of reasoning, there can be multiple quantum jumps of all sorts as the inflaton evolves with time, leading to volumes of space (bubbles) with different inflaton trajectories and, consequently, different cosmological properties. For example, some are flat but some not; some have a scale-invariant spectrum, some not; etc.  

Ultimately, the result is an eternal multiverse in which ``anything can happen and will happen an infinite number of times'' \cite{Guth:2013sya}.
What does inflation predict to be the most likely outcome in the multiverse?  In the context of classical inflation, where volume is the natural measure, most volume today is inflating and most non-inflating volume (bubbles) is predicted to be exponentially younger than the observable universe \cite{Linde:1994gy,Guth:2000ka}.  To be more specific, the volume-weighted prediction \cite{Guth:2013sya} is that our observable universe is exponentially unlikely by a factor exceeding $10^{-10^{55}}$ or more!    By this measure, classic inflation is a catastrophic failure, numerically one of the worst failures in the history of science.

\section{New problems for  inflation after Planck2015?}

Many theorists and observers ignore the conceptual problems described in the previous section.  Their approach is to assume without justification ideal initial conditions and slow-roll inflation with no eternal inflation and see what the simplest potentials predict.   Until now, they could argue that the  `predictions'  are in agreement with observations.  

Planck2015 changed that:
On the one hand, Planck has shown that the non-gaussianity of the density perturbations is small.  This eliminates a wide spectrum of more complex inflationary models and favors models with a single scalar field.  The restriction to single-field models justifies focusing on the plot of $r$  versus $n_S$, since it is optimally designed to discriminate among the single-field possibilities.  
On the other hand,  Planck2015, combined with WMAP and ACT, disfavors all simple single-field models of inflation.  We have already noted that the equation-of-state parameter $\epsilon$ must be $< 1$ to have inflation and $\ll 1$ to produce a scale-invariant spectrum.  A further constraint is that this condition must be maintained for many Hubble expansion times.  The simplest way to achieve these conditions is if  $\epsilon$ is small and nearly constant ($d \ln \epsilon/dN \ll \epsilon$) during most of the slow-roll period.   Simple potentials with a single scalar field and a single parameter suffice to obtain this condition; these are the examples presented in textbooks. 

The problem for these models is that they require a large value of $r$ in order to be compatible with the observed scalar spectral tilt; the tensor-to-scalar ratio  depends linearly on $\epsilon$:
\begin{equation} \label{req}
 r \approx 16 \epsilon.
 \end{equation}
Combining Eq.~\eqref{ns} with the condition that $d \ln \epsilon/dN \ll \epsilon$, we obtain $\epsilon \approx 0.015$ and, hence, $r \approx 0.24$. No wonder that some theorists greeted the BICEP2 announcement for $r \approx 0.2$ as proof of inflation!   

\begin{figure}[!tb]
\centering
\includegraphics[width=9cm]{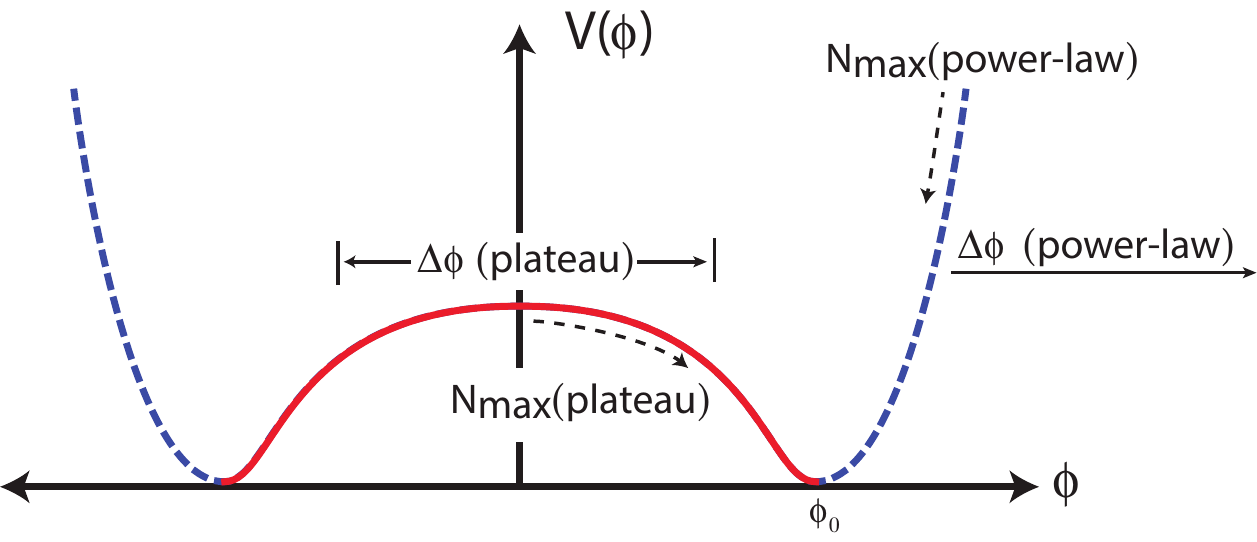}
\caption{Potential $V(\phi)$ allows inflation both on the plateau at small field values, $|\phi| \ll \phi_0$ (solid red), and along the upward power-law curve at large $\phi$, $|\phi |\gg \phi_0$ (dashed blue).  $N_{\text{max}}(\text{plateau/power-law})$ is the maximum number of $e$-folds of inflation possible for the two field ranges $\Delta \phi$, which occurs when $\phi$ slow-rolls along the  trajectories indicated by the dashed curves with arrows. Because $N_{\text{max}}(\text{plateau}) \ll N_{\text{max}}(\text{power-law})$, the fraction of space that undergoes plateau-like inflation  and ends  in the ground state (the minimum of $V$) is exponentially small compared to the fraction of space that undergoes power-law inflation and ends in the same state.  Consequently, an observer in the present universe, which here corresponds to the ground state, is  highly unlikely to have undergone plateau-like inflation.
} 
\label{fig:1}
\end{figure}

\begin{figure}[!tb]
\centering
\includegraphics[width=9cm]{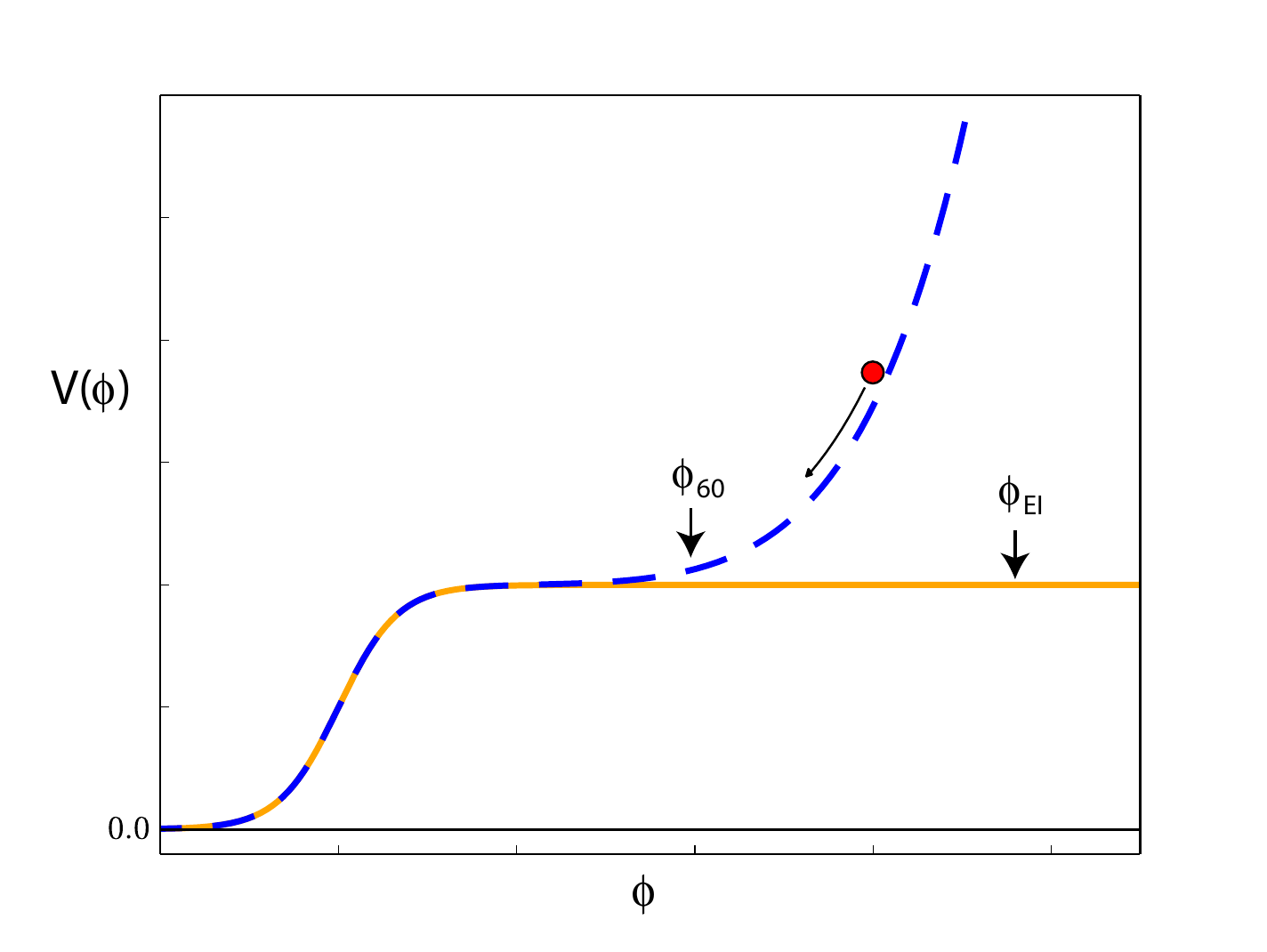}
\caption{For large-field plateau potential (solid curve), patches of space with values of $\phi> \phi_{\rm EI}$ have large density fluctuations ($\delta \rho/\rho >1$) that trigger eternal inflation and the multiverse.  A proposed way to avoid eternal inflation is to add to the potential a wall, a singular  contribution that causes the potential to diverge at some selected value  of $\phi_m> \phi_{\rm EI}$.  The wall causes two problems, though.  First, the cosmological predictions become ultra-sensitive to the initial value of $\phi$ along the wall  and its initial time-variation.  Also, in typical cases, the wall modifies the shape of the plateau near the 60 e-fold mark before the end of inflation ($\phi=\phi_{60}$) enough to shift the predictions for $r$ and $n_S$ so as to violate observational bounds.
} 
\label{fig:2}
\end{figure}

Now that  Planck2015 has re-established that $r< 0.1$,  the entire range of simple models is disfavored and, one might think, inflation should be regarded as disproven.   Attempting to save inflation, theorists have been forced to turn to a set of tuned potentials described as `plateau models'  in which the potential energy density asymptotes at small or large values of the inflaton field to a maximum value of potential energy density that is several orders below the Planck scale; see Figs.~\ref{fig:1} and~\ref{fig:2}. The common feature of the surviving models is that the equation of state satisfies a delicate, unstable condition, $d \ln \epsilon/dN \gg \epsilon$;  that is, the time-variation of $\epsilon$ is large compared to $\epsilon$.  Hence, it is not surprising that plateau models require more parameters and more tuning than the simplest textbook models.  

In addition, plateau models introduce new problems for inflation that the simplest models do not share:

{\it The unlikeliness problem:}
The unlikeliness problem \cite{Ijjas:2013vea} arises because plateau-like potentials require more tuning, occur for a narrower range of initial conditions, and produce exponentially less inflation than would be produced by the now-observationally disfavored power-law potentials,  so it is worrisome to find plateau models to be the only types allowed by the data. Furthermore, most energy landscapes with plateau-like inflation paths to the current vacuum also include simple power-law inflation paths to the same vacuum that generate more inflation, as shown in Fig.~(\ref{fig:1}), so it is exponentially unlikely that the current vacuum resulted from the plateau-like path. Yet this is what is favored after Planck2015.

{\it New initial conditions problem:}
A new initial conditions problem arises because, compared to the simplest inflaton potentials, the energy density at the beginning of inflation $(M_b)^4$ is smaller by twelve orders of magnitude in the observationally favored models.  In order for inflation to begin, a smooth patch of size $(M_b)^{-3}$ Hubble volumes (as evaluated at the Planck time in Planck units) is required.  Quantitatively, the observationally favored potentials require an initial smooth patch that is typically $10^9$ Hubble volumes -- a billion times larger than what is needed to begin inflation for the simplest inflaton potentials (where our numerics assume typical patches dominated by kinetic and radiation energy density).  Since larger smooth patches are exponentially rare than smaller ones, the plateau potentials require initial conditions that are highly improbably compared to the conditions required for the simplest inflaton potentials. 

{\it Exacerbated multiverse problem:}
For classic inflation, volume-weighting was considered fine for making predictions until the discovery of the multiverse, when it was found that Hubble-sized patches of space like ours are highly improbable in the multiverse.  The challenge for the last three decades has been to find an alternative weighting in the multiverse that will restore the naive volume-weighted predictions.   That program has been unsuccessful to date, so there is no justification for expecting that a  plateau potential should produce values of $n_S$, $r$ and $f_{\textsc{nl}}$ that agree precisely with the naive volume-weighted predictions; yet these are the values that Planck2013 and Planck2015 have found. 
This imposes a new tight constraint on any solution to the measure problem:  one must seek a clever choice of weighting that can reproduce the naive volume-weighted predictions of classic inflation for plateau-potentials.  However, then there is another twist.  Using the same naive volume-weighting,  simple potentials are exponentially favored over the small-field plateau models.  Hence, the solution to the measure problem must mimic naive volume-weighting for some predictions but not for others.  These are new data-imposed restrictions for solving the measure problem. 

Mukhanov has  recently argued that the multiverse problem can be evaded in plateau models by terminating the plateau with a steep wall in which the potential energy density of the inflaton field $\phi$ diverges as $1/(\phi- \phi_0)^n$ for some constant $\phi_0$ and some integer $n \ge 4$ \cite{Mukhanov:2014uwa}.  As illustrated in Fig.~\ref{fig:2}, the wall adds an upward twist to the inflaton potential where the value of $\phi_0$ is chosen such that the wall  allows a long enough plateau to have more than 60 $e$-folds of inflation and, at the same time, blocks the inflaton from reaching the large field values where large quantum fluctuations initiate eternal inflation.  Unfortunately, this approach introduces other problems that are just as bad as the multiverse problem.  Not only is the construction awkward, but the wall makes the outcome for the perturbation spectrum ultra-sensitive to the initial conditions of the inflaton field and other forms of energy density.  Furthermore,  because the predictions of a plateau model are ultra-sensitive to the parameters that govern its extent and flatness,  the addition of a wall to a plateau potential will often shift the predictions for $r$ and $n_S$ in a way that violates observational bounds \cite{PresleyXXX}. 

\section{Postmodern inflation}

Some theorists \cite{Guth:2013sya,Linde:2014nna} have argued that the problems of classic inflation, both the longstanding conceptual issues of fine-tuning, initial conditions, and multiverse   and the new issues raised by Planck2015 and other recent cosmic microwave background measurements, are irrelevant.  They advocate abandoning the classic inflationary picture in favor of a different paradigm that we have dubbed {\it postmodern inflation} \cite{Ijjas:2014nta} to emphasize that, while there remains a period of accelerated expansion, the postmodern picture is otherwise a radical departure from the classic one.

In particular, compared to the classic inflationary picture, the postmodern approach makes fundamentally different assumptions about the three inputs needed to make inflationary predictions \cite{Guth:2013sya,Linde:2014nna}:
\begin{itemize}[label=--]  
\item Instead of aiming for simple inflaton potentials, highly complex potentials with many parameters, tunings, and fields are considered ``very plausible according to recent ideas in high-energy physics'' \cite{Guth:2013sya}. 
The complex potentials tend to lead to multiple stages of inflation and a multiverse in which any outcome can occur.  
\item The validity of the postmodern inflationary paradigm cannot be judged on whether it works for typical initial conditions, it is argued, since we do not know what those conditions are.  
\item The volume measure is rejected in favor of complex measures that are to be adjusted ({\it a posteriori}) such that the  likely outcome based on the adjusted measure is guaranteed to agree with observations. 
\end{itemize}
Postmodern inflation has its own issues:

\noindent
{\it Parameter unpredictability problem:}
 Even if the initial conditions were somehow fixed ideally and the multiverse avoided, complex potentials introduce their own form of unpredictability. For example, it has been shown \cite{Ferrara:2013rsa} that a potential with a single field and only three parameters can be designed to fit any outcome for the standard cosmological observables. Another example is the recent spate of $\alpha$-attractor models; although they are described as "attractors," they have adjustable parameters that allow any outcome for $r$ \cite{Kallosh:2013hoa,Kallosh:2013maa,Kallosh:2013daa,Kallosh:2013tua,Kallosh:2013yoa,Kallosh:2014laa,Ferrara:2014kva,Galante:2014ifa,Kallosh:2015lwa,Kallosh:2015zsa,Linde:2015uga,Carrasco:2015uma,Carrasco:2015rva,Carrasco:2015pla}.  With such freedom,  no observation can be said to test the theory. Introducing more degrees of freedom or a complex energy landscape further exacerbates the situation.
 
\noindent
{\it Inflation does not solve the horizon and flatness problems:}
The postmodern view  -- that it is unimportant whether inflation works for typical initial conditions -- is a major shift from the conventional view.   The common justification for introducing inflation is to explain how the observed universe can naturally emerge from a wide range of possible big bang initial conditions. Inflation was supposed to accomplish this by transforming typical initial conditions after the big bang into the smooth and flat conditions observed today.    Several groups have explored the dependence on initial conditions, with some ultimately concluding that the conditions required to have a long period of classic inflation after the universe emerges from the big bang are extremely rare \cite{Penrose:1988mg,Gibbons:2006pa}.  In postmodern inflation, it is conceded that the period of rapid accelerated expansion by itself does not explain how the universe emerged from typical initial conditions.  Ignorance of initial conditions is claimed instead, and the resolution for how the current universe emerged from initial conditions is relegated to the {\it a posteriori} measure.  This means  inflation per se is no longer responsible for resolving the horizon and flatness problems, even though these are the issues  that it was originally introduced to resolve.

Postmodern inflation rests entirely on the measure. It is the measure alone that is supposed to justify the choice of a particular highly complex  potential among exceedingly many. At the same time, the measure is supposed to solve the initial conditions problem, and the very same measure is supposed to regulate infinities in the multiverse and restore predictiveness.  

Such a measure does not currently exist.
Common-sense volume-weighting of classic inflation is declared invalid.   There is no mathematical or logical or intuitive  inconsistency with using the volume measure. In fact, the volume measure may work well for some alternative cosmologies \cite{Johnson:2011aa}. Rather, the only reason that volume-weighting is discarded in postmodern inflation because it leads to a catastrophic prediction:   the observed universe is exponentially unlikely based on volume-weighting. 

\section{Ekpyrotic bouncing cosmology after Planck2015}

Unlike the case of inflation,
ekpyrotic cosmology has not been forced towards more contrived versions as a result of Planck 2015. In fact, it has been recently shown that the simplest version, {\it i.e.}, models that require the least number of parameters and the least amount of tuning, fit the Planck2015 data well.

 Ekpyrotic cosmology \cite{Khoury:2001bz} explains the smoothness and flatness of the cosmological background as being due to  a period of ultraslow contraction before the big bang.  Smoothing during contraction rather than inflation has two significant advantages: it does not require improbable initial conditions and it avoids the multiverse problem.

To have smoothing contraction, the ekpyrotic energy density has to dominate all other forms of stress-energy for an extended period of time. In an approximately Friedmann-Robertson-Walker (FRW) universe with space-time metric $ds^2 = - dt^2 + a^2(t)dx_i dx^i$ where $t$ is the FRW time and $a(t)$ is the scale factor, this means that in the Friedmann equation
\begin{equation}
\label{Friedmannekp}
3M_{\rm Pl}^2H^2 = \frac{\rho_S^0}{a^{2\epsilon}} - \frac{3k}{a^2} + \frac{\sigma^2_0}{a^6}  + [{\rm matter, radiation, etc.}]
\end{equation}
the energy density in the ekpyrotic smoothing component ($\rho_S = \rho_S^0/a^{2 \epsilon}$ where $\rho_S^0$ is a constant) overtakes all other forms of
energy density, including matter ($\rho \propto a^{-3}$), radiation ($\rho \propto a^{-4}$), and gradient energy ($\rho \propto a^{-2}$), and also overtakes the anisotropy ($ \sigma_0^2/ a^6$) and spatial curvature ($k/a^2$). 
Here, $M_{\rm Pl}^2 = (8\pi{\rm G})^{-1}$ is the reduced Planck mass, ${\rm G}$ is Newton's constant, $\epsilon=\rho_S/p_S$ is the equation-of-state parameter of the ekpyrotic energy density, $H=d\ln a/dt$ is the Hubble parameter and dot denotes differentiation with respect to $t$. Hence, the simple condition for ekpyrotic smoothing is $\epsilon>3$.

Currently, the best understood way of creating primordial density perturbations in an ekpyrotic phase is the entropic mechanism \cite{Lehners:2007ac,Buchbinder:2007ad}. Here, pre-bang isocurvature fluctuations are generated by adding a second field. The isocurvature modes are then converted into density perturbations that source structure in the post-bang universe. Note that the ekpyrotic tensor amplitudes are exponentially suppressed  and can be considered negligible \cite{Boyle:2003km,Baumann:2007zm}. 

A simple example of an action describing the standard ekpyrotic mechanism is
\begin{eqnarray}
\label{old_action}
S &=& \int d^4 x \sqrt{-g}\frac{1}{2}R\nonumber\\
&-& \int d^4 x \sqrt{-g}\bigg(\frac{1}{2}\partial_{\mu}\phi_1\partial^{\mu}\phi_1+V_1 e^{-c_1\phi_1}\bigg) 
\nonumber\\
&-&\int d^4 x \sqrt{-g}\bigg(\frac{1}{2}\partial_{\mu}\phi_2\partial^{\mu}\phi_2 + V_2 e^{-c_2\phi_2}\bigg),
\end{eqnarray}
where $V_1, V_2, c_1, c_2$ are constants and the two fields $\phi_1, \phi_2$ have separate ekpyrotic potentials. The background evolution is determined by the linear combination of these potentials while the evolution of perturbations is governed by the entropy field, which is, by definition, perpendicular to the background field. At the end of the ekpyrotic phase and before the bounce, the background trajectory bends and  the isocurvature perturbations are converted into adiabatic ones.
However, it is well-known that these ekyprotic solutions for $\phi_1$ and $\phi_2$ are {\it unstable}, in that the background direction runs along a ridge in the potential that is unstable to variations in the entropic direction.
In addition, to obtain nearly scale-invariant spectra requires a steep negative potential which results in the generation of non-negligible non-gaussianity {\it during the ekpyrotic phase} that dominates the non-gaussianity generated during the conversion of entropic fluctuations to curvature fluctuations after the ekpyrotic phase \cite{Koyama:2007if,Buchbinder:2007at,Lehners:2007wc,Lehners:2008my}.
The steepness of the potential and the instability involve additional tuning of parameters and initial conditions.

Recently, a new type of ekpyrotic mechanism has been discovered that evades these problems and, at the same time, fits the Planck2015 data \cite{Li:2013hga,Fertig:2013kwa,Ijjas:2014fja}. As can be seen by the defining action,
\begin{eqnarray}
\label{action}
S &=& \int d^4 x \sqrt{-g}\frac{1}{2} R\\
&-& \int d^4 x \sqrt{-g}\bigg(\frac{1}{2}\partial_{\mu}\phi\partial^{\mu}\phi + V(\phi) +  \frac{1}{2}\Omega^2(\phi)\partial_{\mu}\chi\partial^{\mu}\chi\bigg),\nonumber
\end{eqnarray}
the new mechanism involves two scalar fields, as before, but only one of the fields has a negative potential, $V(\phi)$.  This first field, $\phi$, dominates the energy density and is the source of the ekpyrotic equation of state.
The crucial ingredient of the new model is a non-trivial field-space metric combined with negligible mass of the second field $\chi$: the non-canonical kinetic coupling $\Omega^2(\phi)$ acts as an additional friction term, ``freezing'' the $\chi$ field. 
Having no or negligible potential, the $\chi$ direction is automatically perpendicular to the background $\phi$ direction in scalar field space. The $\dot{\chi} = 0$ solution naturally defines $\chi$ as the entropy field generating first-order entropy/isocurvature fluctuations while $\phi$ remains the adiabatic field controlling the background evolution. By a standard stability analysis, it can easily be shown that the scale-invariant $(\Omega^2, V)$ solutions for $\phi$ and $H$  are stable (see also \cite{Levy:2015awa}).

Furthermore, as realized in \cite{Ijjas:2014fja}, scale-invariant entropic perturbations can be produced continuously as modes leave the horizon for {\it any time-dependent} ekpyrotic background equation of state. This has the additional advantage of reducing fine-tuning constraints. The corresponding background solutions are stable and the bispectrum of these perturbations vanishes, such that no non-gaussianity is produced during the ekpyrotic phase. Hence, the only contribution to non-gaussianity comes from the non-linearity of the conversion process during which entropic perturbations are turned into adiabatic ones, and this is an ${\cal O}(1)$ contribution to $f_{NL}$.  By these measures, these models are not only the simplest known versions of the ekpyrotic scenario in terms of the degrees of freedom and tuning and, at the same time, fit well within the Planck2015 bounds on non-gaussianity and fit with all other current observational constraints.

To complete the ekpyrotic picture, one has to consider the initial conditions needed to begin the slow contraction phase, the bounce that ends it, and how the combination fits into a full  cosmological history.   
Perhaps the most intriguing possibility for the full history is that the bounces repeat at regular intervals every trillion years or so, such that the evolution of the universe is cyclic \cite{Steinhardt:2002ih}.  At the classical level, the ekpyrotic cyclic model, when combined with the Higgs mechanism, can be made geodesically complete \cite{Bars:2013vba,Bars:2013qna}, the only known example of a geodesically complete theory that incorporates the phases of radiation and matter domination know to have occurred in the visible universe. 

Unlike inflation, the ekpyrotic phase begins when the universe is large and well-described by predictable semi-classical physics, much like the present phase of cosmic acceleration.  An empty volume that is roughly a meter across at that time suffices as an initial condition; it will evolve into a patch the size of the observable universe today after taking account the accelerated expansion, slow contraction, bounce, reheating, and expansion to the present \cite{Erickson:2006wc}.  We know such volumes are abundant today, so they were also abundant a cycle ago.  

For the bounce, two types of possibilities are being explored:  `non-singular' bounces in which the scale factor $a(t)$ shrinks to a finite value and rebounds; and `singular' bounces in which the scale factor shrinks to zero and then rebounds.  The non-singular bounce requires violation of the null-energy condition or modifications to Einstein gravity \cite{Buchbinder:2007ad,Rubakov:2014jja}.  An appealing feature of the non-singular bounce is that the universe never reaches high densities where quantum gravity and UV completion are important.  
The singular bounce must take account of quantum gravity effects near the bounce.  Difficult as this might seem, there are several lines of argument \cite{Bars:2013vba,Bars:2013qna,Gielen:2015uaa,Bars:2015trh} that suggest  quantum gravity may not affect the bounce or the properties of the universe at long-wavelengths compared to the Planck scale except for passing through the curvature perturbations that were created during the ekpyrotic phase.   Furthering and making more rigorous our understanding of the bounce is an active and critical area of investigation.

\section{Anamorphic cosmology}

Most recently, a novel approach to the early-universe has been proposed. Anamorphic cosmology \cite{Ijjas:2015zma} combines advantages of inflationary and ekpyrotic models of the early universe without their disadvantages. 
It is based on the assumption that, during the primordial genesis phase, the Planck mass $M_{\rm Pl}$ and the mass of massive particles $m$ have different time dependence in any Weyl frame. (For simplicity, we consider  models in which matter-radiation consists of massive dust and the action for a single particle is $S_p=\int m ds$, where $ds$ is the line element and $m$ may vary with time.) This simple amendment to Einstein's general relativity leads to interesting consequences: most importantly, no absolute notion about expansion and contraction of the cosmological background exists in such theories.   Instead, the background behavior, whether it is expanding or contracting, must now be characterized relative to the evolution of a test particle or the Planck mass. 

Consequently, while the Hubble parameter $H=\dot{a}/a$ is a good quantity to fully capture the background behavior of a homogeneous and isotropic FRW universe in conventional general relativity, 
two new independent quantities are required to characterize the background in anamorphic cosmology: one to characterize the background behavior relative to the characteristic matter scale ({\it e.g.}, Compton wavelength) and one to characterize the background relative to the gravitational (Planck) scale. 
A dimensionless quantity that describes the physical expansion or contraction of the cosmological background
as measured relative to a ruler (or any object made of matter) is given by
\begin{equation}
\Theta_m = \left( H + \frac{\dot{m}}{m} \right)
.
\end{equation}
The corresponding dimensionless quantity that measures the evolution
relative to the Planck mass, which is important for analyzing the generation of scalar and tensor metric perturbations is given by
\begin{equation}
\Theta_{\rm Pl} = \left( H + \frac{\dot{M}_{\rm Pl}}{M_{\rm Pl}} \right)
.
\end{equation}
The variables $\Theta_{\rm Pl}$ and $\Theta_m$ are particularly useful since they are Weyl-frame independent. In ordinary FRW cosmology, $\Theta_{\rm Pl}$ and $\Theta_m$ have the same sign and are both equal to $H$.

The defining feature of anamorphic cosmology is that the Hubble-like parameters $\Theta_{\rm Pl}$ and $\Theta_m$ have opposite signs: $\Theta_m$ is negative and $\Theta_{\rm Pl}$ is positive. Such a model can be realized, for example, in scalar-tensor theories of gravity as has been shown in \cite{Ijjas:2015zma}.

During the anamorphic phase, it is most useful to express 
the first Friedmann equation  in a frame-invariant form using $\Theta_m$:
\begin{eqnarray}
\label{FriedmannEq1}
3\,  \Theta_m^2 \left(1 -  \frac{ d\ln \left( m/M_{\rm Pl} \right) }{ d\ln \alpha_m} \right)^2
= \frac{ \rho_{\rm A} }{ M_{\rm Pl}^4 }
&+& \frac{\rho_m}{M_{\rm Pl}^4 }
- \left(\frac{m}{M_{\rm Pl}}\right)^2 \frac{\kappa}{\alpha_m^2} \nonumber \\
&+ & \left(\frac{m}{ M_{\rm Pl} }\right)^6 \frac{\sigma^2}{\alpha_m^6},
\end{eqnarray}
where the effective scale factor is given through $ \Theta_m \equiv M_{\rm Pl}^{-1}( \dot{\alpha}_m/\alpha_m) $.
The Friedmann equation describes the contributions of different forms of energy density  and curvature to the contraction or expansion rate.   The first contribution, $\rho_{\rm A}/M_{\rm Pl}^4 $, is due to the anamorphic energy density that dominates during the smoothing phase. The second contribution, $\rho_{\rm matter}/M_{\rm Pl}^4$, is due to the matter-radiation energy density in which the matter consists of particles with mass $m$.  The last two contributions are due to the spatial curvature, where $\kappa = (+1,0,-1)$,  and to the anisotropy, parameterized by $\sigma^2$.  

The anamorphic combination of $\Theta_m < 0 < \Theta_{\rm Pl}$ requires that $m$ and/or $M_{\rm Pl}$ be time-dependent such that the invariant mass ratio $m/ M_{\rm Pl}$ decreases with time.  
Note that, under this condition, the factors of $m/M_{\rm Pl}$ in the Friedmann equation suppress the spatial curvature and, even more so, the anisotropy. This is a key feature because suppressing the anisotropy in a contracting universe is essential for avoiding chaotic mixmaster behavior and maintaining homogeneity and isotropy.    
At some point after the anamorphic phase, to reach consistency with all current cosmological observations and tests of general relativity,
the universe must reheat to a high temperature and enter a hot expanding phase in which  both $m$ and $M_{\rm Pl}$  become constant, with  $M_{\rm Pl}=M_{\rm Pl}^0$,  the current value of the reduced Planck mass.
Here, we use reduced Planck units with $M_{\rm Pl}^0\equiv 1$, except where specified otherwise.

To resolve the homogeneity and isotropy problems and to generate nearly scale-invariant perturbations, the anamorphic energy density must dominate all other contributions on the right hand side of the Friedmann equation~\eqref{FriedmannEq1} for a sufficiently long time ($\sim 60$ $e$-folds). This means, 
\begin{equation}
\label{smoothing}
\Theta_m^2 (1-q)^2 \propto  \rho_{\rm A}/M_{\rm Pl}^4 \propto 1/ \alpha_m^{ 2 \epsilon_m },
\end{equation}
 where $q$ is the mass-variation index given by $1-q = d\alpha_{\rm Pl}/d\alpha_m$ and $\epsilon_m$ is the equation-of-state parameter. Note that $q>1$ because $\Theta_m<0<\Theta_{\rm Pl}$.
In order for Eq.~\eqref{smoothing} to hold as $\alpha_m$ shrinks during the anamorphic contracting phase, $2\epsilon_m $ must exceed the corresponding exponents for the spatial curvature and anisotropy terms in the Friedmann equation~\eqref{FriedmannEq1} if they are expressed as powers of $1/\alpha_m$.  
This condition yields a pair of constraints on $\epsilon_m$:
\begin{eqnarray}
\label{smcon}
 \epsilon_m   \gtrsim 1 - q \quad \& \quad \epsilon_m   \gtrsim 3(1-q). 
\end{eqnarray}
Here, for simplicity, we have neglected the weak time-dependence of $q$. 
Since $q > 1$, both conditions are satisfied if the first inequality is satisfied, i.e., if $\epsilon_m \gtrsim 1-q$.   
Note that, because $\Theta_m$ is negative,
the cosmological background is physically contracting, so the smoothing occurs without creating a multiverse or incurring an initial conditions problem, as in ekpyrotic models. 

To obtain a nearly scale-invariant spectrum of super-horizon curvature perturbations, the cosmological background must have the property that modes whose wavelengths are inside the horizon at the beginning of the smoothing phase can have wavelengths larger than the horizon size by the end of the smoothing phase.  The `horizon' is a dynamical length scale that separates smaller wavelengths for which the curvature modes are oscillatory from the large wavelengths for which the curvature modes become frozen.  In anamorphic models, the evolution of scalar and tensor metric perturbations are described  entirely by the gravitational and scalar field parts of the effective action and do not depend on particle mass.  Hence, the dynamical length scale is set by $\Theta_{\rm Pl}^{-1}$ and the corresponding squeezing condition  
is that $\alpha_{\rm Pl} \Theta_{\rm Pl} $ be increasing,  
\begin{equation}
\label{sq1}
\frac{d |\alpha_{\rm Pl}\Theta_{\rm Pl}|}{d\,t} M_{\rm Pl}^{-1}  > 0,
\end{equation}
which reduces to the standard condition in inflationary and ekpyrotic models. The effective scale factor $\alpha_{\rm Pl}$ is given through $ \Theta_{\rm Pl} \equiv M_{\rm Pl}^{-1}( \dot{\alpha}_{\rm Pl}/\alpha_{\rm Pl}) $.
As shown in \cite{Ijjas:2015zma}, 
 the squeezing constraint reduces to the same condition as the smoothing constraint and, 
hence, squeezing imposes no additional constraint.
Because $\Theta_{\rm Pl}$ is positive, the second-order action describing the generation and evolution of curvature perturbations during the smoothing phase is similar to the case of inflation. Consequently, a nearly scale-invariant spectrum of adiabatic density and gravitational wave
perturbations can be generated in models with a single scalar field.
 
It is important to note that anamorphic cosmology, like ekpyrotic cosmology, requires a bounce, {\it i.e.}, a transition from smoothing contraction ($\Theta_m<0$) to standard big-bang expansion ($\Theta_m>0$). The anamorphic picture may potentially be made cyclic and geodesically complete, as well \cite{anamorphic2}.
An anamorphic bounce can be realized more simply, though:  For example, the scalar-tensor realization of anamorphic cosmology allows for the Hubble-like parameter $\Theta_m$ to increase with time and eventually hit zero ($\Theta_m$-bounce) while $\Theta_{\rm Pl}>0$, all without violating the null-energy condition and without instabilities.

\section{Discussion}

Planck2015 calls for a simple theory of cosmic origins.   Observations currently suggest that the large-scale structure of the universe is nearly as scale-free and featureless as quantum physics will allow.  

A notable achievement is that Planck2015, when combined with earlier measurements, provides the first definitive evidence strongly disfavoring the simplest inflationary models.  Many theorists have now turned their focus to the  next-best inflationary alternatives but, as we have explained, these ``plateau'' models introduce several new problems: more parameters, more tuning, more implausible initial conditions, an exacerbated multiverse problem, and the unlikeliness problem.  The added complexity does not fit well with the remarkable simplicity of the large-scale universe indicated by Planck2015 and other observations.

Some have advocated discarding the classic inflationary theory in favor of a postmodern picture.   The postmodern paradigm accepts a multiverse in which anything can happen, with initial conditions yet to be determined,  with complex potentials consisting of multiple fields and parameters, and, then, with the freedom to select the measure {\it a posteriori}.  Data has no significance for the postmodern paradigm because any observed outcome can be retrofit -- failure to match observations can always be corrected with a change of measure.   Hence, the postmodern view is empirically untestable by construction and, in this sense, a departure from normal science.  

Bouncing models are promising alternatives that can explain Planck2015 and other observations without introducing problems with initial conditions or producing a multiverse. A signficant recent advance described here is the discovery of a simplified type of ekpyrotic bouncing cosmology that does not produce large non-gaussianity or instabilities like previous versions did and that requires less tuning.   Like all ekpyrotic models, the amplitude of the tensor mode is far too small to be observed in any plausible measurements of cosmic microwave background polarization.  

An important parallel development is the anamorphic scenario, which introduces modifications to Einstein's general relativity theory to produce cosmologies that are, in one sense, expanding and, in other sense, contracting.  The anamorphic or dual picture makes it possible to combine the advantages of inflationary and bouncing cosmology without the disadvantages of either.  In some versions, anamorphic models can produce a detectable spectrum of tensor modes and a measurable value of $r$. 

It will be important to see whether forthcoming measurements of B-mode polarization and non-gaussianity continue to indicate this simple large-scale structure, since this is surely a profound hint about the theory that will ultimately explain the origin, evolution and future of the universe.

{\it Acknowledgments.}
This work was supported in part by the US Department of Energy grant DE-FG02-91ER40671 (PJS).

\bibliographystyle{apsrev}
\bibliography{PlanckInflation}

\end{document}